\documentclass[twocolumn]{jpsj2} 
\newcommand{\iu}{\rm i}
\newcommand{\rmd}{{\rm d}}
\newcommand{\bmath}[1]{\mbox{\boldmath$#1$}}

\title{Mechanism of Anomalous Tunneling in Condensed Bose System}

\author{\textsc{Yusuke Kato}$^{1}$\thanks{E-mail address: yusuke@phys.c.u-tokyo.ac.jp}, \textsc{Hiroshi Nishiwaki}$^{1}$ and \textsc{Akitake Fujita}$^{2}$}

\inst{$^{1}$Department of Basic Science, University of Tokyo, Tokyo 153-8902\\
$^{2}$Department of Physics, University of Tokyo, Tokyo 113-0033\\
}

\abst{
We clarify the origin of anomalous tunneling [Yu. Kagan et al. Phys. Rev. Lett. 90 (2003) 130402] i.e. the perfect transmission at low energy limit of tunneling of phonon excitations across the potential barrier separating two Bose condensates. The perfect transmission is a consequence of the coincidence of the wave function of the excited state at low energy limit and the macroscopic wave function of the condensate. We show that the perfect transmission at low energy occurs even at finite temperatures within the scheme of Popov approximation. 
}

\kword{BEC, Tunneling, Gross-Pitaevskii equation, Bogoliubov equation}

\begin{document}
\maketitle
Bose-Einstein condensation in trapped dilute alkali-metal atoms has opened up a chance to find new quantum phenomena, owing to the controllability of the strength of interaction by Feshbach resonance and of a one-body potential by focused laser beam. With these backgrounds, many theoretical studies have been done with the Gross-Pitaevskii(GP) and Bogoliubov equations\cite{pethicksmith}, which are no longer theoretical toy models but describe weakly interacting Bose gases realized in alkali-metal atoms. One of those theoretical studies has been done by Kovrizhin\cite{Kov} and Kagan et al.\cite{Kagan}. They considered tunneling of excitation over the barrier separating the two condensates and found that the perfect transmission occurs at low energy limit, which was named ^^ ^^ anomalous tunneling''\cite{Kagan}. Subsequently, Danshita et al. \cite{DYK1,DYK2,Danshita,DYK3} considered, extensively, related problems such as tunneling between two condensates with different phases\cite{DYK1,DYK2,DYK3} and the excitation spectrum in Kronig-Penny potential\cite{DKT,Danshita}. The transmission of excitations through randomly distributed potential barriers was discussed by Bilas and Pavloff\cite{BilasPavloff}.  

Kagan et al. attributed the anomalous tunneling to a virtual resonance level at low energy, which results from the repulsive potential barrier and the depletion of condensate wave function near the barrier. Through the study of tunneling between the two condensates with different phases, on the other hand, Danshita et al. \cite{DYK1} found that anomalous tunneling occurs only when the localized component is nonzero in the asymptotic solution of Bogoliubov equation. Earlier arguments \cite{Kagan, DYK1} of the origin of anomalous tunneling, however, rely on the explicit analytical solutions of GP and Bogoliubov equations in the presence of a particular form of potential barrier; $V(x)\propto \delta(x)$ or the rectangular well $V(x)=V_0(>0)$ for $|x|<a$ and 0 otherwise ($x$-axis is taken to be the normal direction of the wall of the potential barrier). Therefore their arguments can not tell whether anomalous tunneling occurs or not in the presence of general form of potential barrier, which does not allow the exact analytical solution for GP and Bogoliubov equations. 
Furthermore, needed is an interpretation applicable to finite temperatures, where the exact analytical solution is not available in general. 

  In this Letter, we show the anomalous tunneling occurs for the potential barrier $V(x)$ being arbitrary but short-ranged and even function $(V(-x)=V(x))$. Further within the Popov approximation, we show the anomalous tunneling occurs even at finite temperatures. 

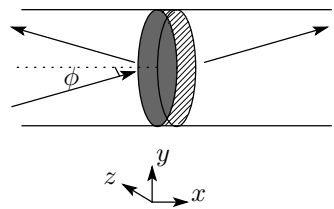
\begin{figure}
\begin{center}
\unitlength 0.1in
\begin{picture}( 16.7100, 10.1300)(  0.0000,-10.1300)
%
\special{pn 8}%
\special{sh 0.600}%
\special{ar 762 306 102 306  0.0000000 6.2831853}%
%
\special{pn 8}%
\special{ar 862 306 100 306  0.0000000 6.2831853}%
%
\special{pn 8}%
\special{pa 762 0}%
\special{pa 762 0}%
\special{fp}%
\special{pa 862 0}%
\special{pa 862 0}%
\special{fp}%
\special{pa 762 0}%
\special{pa 862 0}%
\special{fp}%
%
\special{pn 8}%
\special{pa 52 0}%
\special{pa 1672 0}%
\special{fp}%
\special{pa 762 0}%
\special{pa 862 0}%
\special{fp}%
%
\special{pn 8}%
\special{pa 52 608}%
\special{pa 1672 608}%
\special{fp}%
%
\special{pn 4}%
\special{pa 952 426}%
\special{pa 816 562}%
\special{fp}%
\special{pa 948 460}%
\special{pa 826 582}%
\special{fp}%
\special{pa 936 500}%
\special{pa 840 598}%
\special{fp}%
\special{pa 918 552}%
\special{pa 872 598}%
\special{fp}%
\special{pa 956 390}%
\special{pa 832 516}%
\special{fp}%
\special{pa 962 356}%
\special{pa 846 472}%
\special{fp}%
\special{pa 962 324}%
\special{pa 852 436}%
\special{fp}%
\special{pa 962 294}%
\special{pa 858 400}%
\special{fp}%
\special{pa 956 268}%
\special{pa 862 366}%
\special{fp}%
\special{pa 956 238}%
\special{pa 862 334}%
\special{fp}%
\special{pa 956 208}%
\special{pa 862 306}%
\special{fp}%
\special{pa 952 184}%
\special{pa 862 274}%
\special{fp}%
\special{pa 948 158}%
\special{pa 862 244}%
\special{fp}%
\special{pa 942 134}%
\special{pa 858 218}%
\special{fp}%
\special{pa 936 106}%
\special{pa 858 188}%
\special{fp}%
\special{pa 928 88}%
\special{pa 852 162}%
\special{fp}%
\special{pa 924 62}%
\special{pa 846 136}%
\special{fp}%
\special{pa 906 44}%
\special{pa 836 116}%
\special{fp}%
\special{pa 896 26}%
\special{pa 832 90}%
\special{fp}%
\special{pa 882 12}%
\special{pa 820 72}%
\special{fp}%
\special{pa 852 12}%
\special{pa 816 44}%
\special{fp}%
%
\special{pn 4}%
\special{pa 832 0}%
\special{pa 802 30}%
\special{fp}%
\special{pa 802 0}%
\special{pa 786 16}%
\special{fp}%
%
\special{pn 4}%
\special{pa 826 582}%
\special{pa 802 608}%
\special{fp}%
%
\special{pn 8}%
\special{pa 760 302}%
\special{pa 28 302}%
\special{dt 0.045}%
%
\special{pn 8}%
\special{pa 0 508}%
\special{pa 642 326}%
\special{fp}%
\special{sh 1}%
\special{pa 642 326}%
\special{pa 572 324}%
\special{pa 590 340}%
\special{pa 582 364}%
\special{pa 642 326}%
\special{fp}%
%
\special{pn 8}%
\special{pa 1008 276}%
\special{pa 1650 92}%
\special{fp}%
\special{sh 1}%
\special{pa 1650 92}%
\special{pa 1580 92}%
\special{pa 1598 108}%
\special{pa 1590 130}%
\special{pa 1650 92}%
\special{fp}%
%
\special{pn 8}%
\special{pa 642 276}%
\special{pa 0 92}%
\special{fp}%
\special{sh 1}%
\special{pa 0 92}%
\special{pa 60 130}%
\special{pa 52 108}%
\special{pa 70 92}%
\special{pa 0 92}%
\special{fp}%
%
\special{pn 8}%
\special{pa 540 302}%
\special{pa 564 348}%
\special{fp}%
\put(2.7000,-4.2000){\makebox(0,0)[lb]{$\phi$}}%
%
\special{pn 8}%
\special{pa 734 1008}%
\special{pa 916 1008}%
\special{fp}%
\special{sh 1}%
\special{pa 916 1008}%
\special{pa 850 988}%
\special{pa 864 1008}%
\special{pa 850 1028}%
\special{pa 916 1008}%
\special{fp}%
%
\special{pn 8}%
\special{pa 734 1008}%
\special{pa 734 824}%
\special{fp}%
\special{sh 1}%
\special{pa 734 824}%
\special{pa 714 892}%
\special{pa 734 878}%
\special{pa 754 892}%
\special{pa 734 824}%
\special{fp}%
%
\special{pn 8}%
\special{pa 734 1008}%
\special{pa 582 916}%
\special{fp}%
\special{sh 1}%
\special{pa 582 916}%
\special{pa 630 968}%
\special{pa 628 944}%
\special{pa 650 934}%
\special{pa 582 916}%
\special{fp}%
\put(9.3900,-10.0300){\makebox(0,0)[lb]{$x$}}%
\put(7.5600,-8.2000){\makebox(0,0)[lb]{$y$}}%
\put(4.8100,-9.1200){\makebox(0,0)[lb]{$z$}}%
\end{picture}%
\end{center}
\caption{Set-up of the system. The two condensates are separated by the barrier with the shape of wall. The normal direction is taken to be $x$-axis. The incident angle of an excitation to the wall is denoted by $\phi$.}
\label{f1}
\end{figure}

First we consider weakly interacting Bose system at zero temperature in the presence of potential barrier $V(x)$.  Linear dimensions in the transverse directions  $(y,z)$ are assumed to be large so that the mean-field approximation is valid. The system we consider is schematically described in Fig.~\ref{f1}. We take the incident angle $\phi$ to be arbitrary within the range $[0,\pi/2)$, while $\phi$ has been taken to be zero in earlier studies\cite{Kov,Kagan,DYK1}. 

In the mean-field theory, the macroscopic wave function $\Psi$ satisfies the stationary Gross-Pitaevskii equation 
\begin{equation}
\hat{H}\Psi=0,
\label{eq: st-GP}
\end{equation}
with 
\begin{equation}
\hat{H}=-\frac{\hbar^2\nabla^2}{2m}+V(x)-\mu+U |\Psi|^2. 
\end{equation}
Here $U$ is the coupling constant of two-body short-range interaction and is related to $4\pi \hbar^2 a_0/m$ with the scattering length $a_0$. Other notations are conventional. 
The system is uniform in $y,z$ directions and hence $\Psi$ depends only on $x$. Far away from the barrier $|x|\gg1$, $\Psi$ approaches $\sqrt{\mu/U}$. The spatial variation of $\Psi(x)$ is scaled by the healing length $\xi=\hbar/\sqrt{m\mu}$. Henceforth $\Psi(x)$ is considered to be real. 
In the presence of $V(x)$ and $\Psi(x)$, the two-component wave function $(u(\bmath{r};\varepsilon),v(\bmath{r};\varepsilon))^t$ of the state with excitation energy $\varepsilon$ satisfies the Bogoliubov equation
  \begin{eqnarray}
  \left(
  \begin{array}{cc}
  \hat{H}'& -U \Psi^2(x)\\
  U\Psi^{*2}(x) &-\hat{H}'\\
  \end{array} 
  \right)
  \left(
  \begin{array}{c}
  u(\bmath{r};\varepsilon) \\
  v(\bmath{r};\varepsilon) \\
  \end{array} 
  \right)=\varepsilon
  \left(
  \begin{array}{c}
  u(\bmath{r};\varepsilon) \\
  v(\bmath{r};\varepsilon) \\
  \end{array} 
  \right)\nonumber\\
  \label{eq: Beq3d}
  \end{eqnarray}
  with $\hat{H}'=\hat{H}+U|\Psi(x)|^2$.   
Owing to the translational invariance in $y,z$ directions, the solution of (\ref{eq: Beq3d}) has the form of 
\begin{equation}
  \left(
  \begin{array}{c}
  u(\bmath{r};\varepsilon) \\
  v(\bmath{r};\varepsilon) \\
  \end{array} 
  \right)=\exp({\iu} k_y y +{\iu} k_z z)
   \left(
  \begin{array}{c}
  u(x;\varepsilon) \\
  v(x;\varepsilon) \\
  \end{array} 
  \right).
\end{equation}
The wave function $(u(x;\varepsilon),v(x;\varepsilon))^{\rm t}$ satisfies the following equation \begin{eqnarray}
  \left(
  \begin{array}{cc}
  \hat{H}''& -U \Psi^2(x)\\
  U\Psi^{*2}(x) &-\hat{H}''\\
  \end{array} 
  \right)
  \left(
  \begin{array}{c}
  u(x;\varepsilon) \\
  v(x;\varepsilon) \\
  \end{array} 
  \right)=\varepsilon
  \left(
  \begin{array}{c}
  u(x;\varepsilon) \\
  v(x;\varepsilon) \\
  \end{array} 
  \right)\nonumber\\
  \label{eq: Beq}
  \end{eqnarray}
  with    
$$\hat{H}''=-\frac{\hbar^2}{2m}\frac{{\rm d}^2}{{\rm d}x^2}+\frac{\hbar^2 k^2_\perp}{2m}+V(x)-\mu+2U |\Psi|^2$$ and 
$k_\perp =\sqrt{k_y^2 +k_z^2}$.

Introducing reduced variables
$\bar{x}=x/\xi$, $\bar{\Psi}({\bar{x}})=\sqrt{U/\mu}\Psi(x)$, $\bar{V}=V/\mu$, ${\bar k}_\perp=k_\perp\xi$ and $\bar{\varepsilon}=\varepsilon/\mu$, equations (\ref{eq: st-GP}) and (\ref{eq: Beq}) reduce to the following equations:
\begin{equation}
\hat{h}\bar{\Psi}(\bar{x})=0
\label{eq: st-GP-red}
\end{equation}
with $\hat{h}=-\frac{1}{2}\frac{\rmd^2}{\rmd \bar{x}^2}+\bar{V}(\bar{x})-1+\bar{\Psi}^2(\bar{x})$ and
 \begin{eqnarray}
  \left(
  \begin{array}{cc}
  \hat{h}''& -\bar{\Psi}^2(\bar{x})\\
  \bar{\Psi}^{2}(\bar{x}) &-\hat{h}''\\
  \end{array} 
  \right)
  \left(
  \begin{array}{c}
  u(\bar{x};\bar{\varepsilon}) \\
  v(\bar{x};\bar{\varepsilon}) \\
  \end{array} 
  \right)
  =\bar{\varepsilon}
  \left(
  \begin{array}{c}
  u(\bar{x};\bar{\varepsilon}) \\
  v(\bar{x};\bar{\varepsilon}) \\
  \end{array} 
  \right),\label{eq: B-eq-red}
  \end{eqnarray}
  with $\hat{h}''=\hat{h}+\bar{\Psi}^2(\bar{x})+\bar{k}^2_\perp/2$, respectively.   
  Henceforth, we omit the bar for simplicity. 
  At $|x|\gg1$, where $V\sim 0$ and $\Psi(x)\sim 1$, the basis of the solution of (\ref{eq: B-eq-red}) is given by the two plane-wave solutions   \begin{equation}
e^{\pm {\rm i} k_\parallel x}  \left(
  \begin{array}{c}
  1 \\
  \theta_+(\varepsilon) \\
  \end{array} 
  \right),
  \label{eq: basis12}
\end{equation}
and exponentially growing or converging solutions
\begin{equation}
  e^{\pm \kappa x}  \left(
  \begin{array}{c}
  1 \\
  \theta_-(\varepsilon) \\
  \end{array} 
  \right),
  \label{eq: basis34}
\end{equation}
with $\theta_\pm(\varepsilon)=\pm \sqrt{1+\varepsilon^2}-\varepsilon$. 
The wave number $k_\parallel$ in the plane-wave solutions (\ref{eq: basis12}) is given by
\begin{equation}
k_\parallel=\sqrt{k^2-k^2_\perp},\quad \mbox{with }k=\sqrt{2(\sqrt{1+\varepsilon^2}-1)}.
\end{equation}
It is convenient to express 
\begin{equation}
k_\parallel=k\cos\phi,\quad k_\perp=k\sin\phi,\quad \phi\in[0,\pi/2)
\end{equation}
in terms of the incident angle $\phi$ with respect to the wall of the potential barrier (see Fig.~1).

The growing rate (or converging rate) $\kappa$ in (\ref{eq: basis34}) is given by
\begin{equation}
\kappa=\sqrt{2(\sqrt{1+\varepsilon^2}+1)+k^2\sin^2\phi}. 
\end{equation}
 
In (\ref{eq: basis34}), the exponentially-diverging component $e^{\kappa|x|}$ should be absent in the physical solution for the wave function. 
When the solution of (\ref{eq: B-eq-red}) has the asymptotic form:
\begin{equation}
(e^{{\rm i} k_\parallel x}+r(\varepsilon) e^{-{\rm i} k_\parallel x})  \left(
  \begin{array}{c}
  1 \\
  \theta_+(\varepsilon) \\
  \end{array} 
  \right)+b(\varepsilon)
e^{\kappa x}  \left(
  \begin{array}{c}
  1 \\
  \theta_-(\varepsilon) \\
  \end{array} 
  \right)
\label{eq: asym-left}
\end{equation}
for $x\ll -1$ and 
\begin{equation}
t(\varepsilon) e^{{\rm i} k_\parallel x}\left(
  \begin{array}{c}
  1 \\
  \theta_+(\varepsilon) \\
  \end{array} 
  \right)+c(\varepsilon)
e^{-\kappa x}  \left(
  \begin{array}{c}
  1 \\
  \theta_-(\varepsilon) \\
  \end{array} 
  \right)
\label{eq: asym-right}
\end{equation}
for $x\gg 1$, 
$t(\varepsilon)$ and $r(\varepsilon)$ are the transmission and reflection coefficients of excitation with energy $\varepsilon$ for the potential barrier $V$. 

Now we derive the solution 
of (\ref{eq: B-eq-red}) corresponding to (\ref{eq: asym-left}) and (\ref{eq: asym-right}) in another way. In terms of 
$$S(x;\varepsilon)=u(x;\varepsilon)+v(x;\varepsilon)$$
and $$G(x;\varepsilon)=u(x;\varepsilon)-v(x;\varepsilon),$$ 
(\ref{eq: B-eq-red}) becomes
\begin{equation}
(\hat{h}+k^2_\perp/2)S(x;\varepsilon)=\varepsilon G(x;\varepsilon),
\end{equation}
\begin{equation}
(\hat{h}^{(-)}+k^2_\perp/2)G(x;\varepsilon)=\varepsilon S(x;\varepsilon),
\label{eq: SG}
\end{equation}
with $\hat{h}^{(-)}=\hat{h}+2\Psi^2(x)$. 
Expanding $S(x;\varepsilon)$ and $G(x;\varepsilon)$ as 
\begin{equation}
S(x;\varepsilon)=\sum_{n=0}^\infty \varepsilon^n S^{(n)}(x),
\quad
G(x;\varepsilon)=\sum_{n=0}^\infty \varepsilon^n G^{(n)}(x)
\end{equation}
with respect to $\varepsilon$, the equations for $S^{(0)}(x)$ $G^{(0)}(x)$ and $S^{(1)}(x)$ are obtained as
\begin{equation}
\hat{h}S^{(0)}(x)=0,
\quad
\hat{h}^{(-)}G^{(0)}(x)=0,
\label{eq: SG-zero}
\end{equation}
and 
\begin{equation}
\hat{h}S^{(1)}(x)=G^{(0)}(x).
\label{eq: S-one}
\end{equation}
Equations (\ref{eq: SG-zero}) and (\ref{eq: S-one}) are sufficient to derive the low energy properties of tunneling of excitations over the potential barrier. 

The solution of the first equation of (\ref{eq: SG-zero}) can be written as the linear combination 
\begin{equation}
S^{(0)}(x)=A^{(0)}_{\rm e}s_{\rm e}(x)+A^{(0)}_{\rm o}s_{\rm o}(x)
\end{equation}
of the even function $s_{\rm e}(x)$ and odd function $s_{\rm o}(x)$ satisfying 
\begin{equation}
\hat{h}s_{\rm e}(x)=0,\quad s_{\rm e}(0)=1,\quad \frac{\rmd s_{\rm e}(x)}{\rmd x}\Big|_{x=0}=0
\label{eq: s-0-e}
\end{equation}
and 
\begin{equation}
\hat{h}s_{\rm o}(x)=0,\quad s_{\rm o}(0)=0,\quad \frac{\rmd s_{\rm o}(x)}{\rmd x}\Big|_{x=0}=1,
\label{eq: s-0-o}
\end{equation}
respectively. At $|x|\gg 1$, where $V(x)\sim 0$ and $\Psi(x)\sim 1$,  $\hat{h}$ reduces to $-\frac12\frac{\rmd^2}{\rmd x^2}$ and as a result, $s_{\rm e}(x)$ and $s_{\rm o}(x)$ have the forms of 
\begin{equation}
s_{\rm e}(x)\sim \alpha_{\rm e}+\beta_{\rm e}|x|,\quad
s_{\rm o}(x)\sim \alpha_{\rm o}{\rm sgn}(x)+\beta_{\rm o}x,
\end{equation}
respectively. Generally, $\alpha_{\rm e}$, $\alpha_{\rm o}$ and $\beta_{\rm o}$ are nonzero; we have confirmed that this is true for $V(x)$ being the rectangular well\cite{Kagan} analytically and $V(x)=\exp(-x^2)$ numerically. On the other hand, the relation 
\begin{equation}
\beta_{\rm e}=0
\label{eq: beta-e-zero}
\end{equation}
holds as a result of   
\begin{equation}
s_{\rm e}(x)=\Psi(x)/\Psi(0),
\label{eq: s-0-e-Psi}
\end{equation}
which follows from the fact that both hand sides of (\ref{eq: s-0-e-Psi}) satisfy the same equation and initial condition (\ref{eq: s-0-e}) and $\Psi(x)/\Psi(0)$ approaches a constant at $|x|\gg 1$. For the moment, we leave the constant $\beta_{\rm e}$ in the expressions of physical quantities in order to make the role of (\ref{eq: beta-e-zero}) manifest in the perfect transmission at low energy limit. 
 
The solution of the second equation of (\ref{eq: SG-zero}) can be written as the linear combination 
\begin{equation}
G^{(0)}(x)=B^{(0)}_{\rm e}g_{\rm e}(x)+B^{(0)}_{\rm o}g_{\rm o}(x)
\label{eq: G-zero-l-c}
\end{equation}
of the even function $g_{\rm e}(x)$ and odd function $g_{\rm o}(x)$ satisfying 
\begin{equation}
\hat{h}^{(-)} g_{\rm e}(x)=0,\quad g_{\rm e}(0)=1,\quad \frac{\rmd g_{\rm e}(x)}{\rmd x}\Big|_{x=0}=0
\label{eq: g-0-e}
\end{equation}
and 
\begin{equation}
\hat{h}^{(-)} g_{\rm o}(x)=0,\quad g_{\rm o}(0)=0,\quad \frac{\rmd g_{\rm o}(x)}{\rmd x}\Big|_{x=0}=1,
\label{eq: g-0-o}
\end{equation}
respectively. At $|x|\gg 1$, where $\hat{h}^{(-)}$ reduces to $-\frac12\frac{\rmd^2}{\rmd x^2}+2$, $g_{\rm e}(x)$ and $g_{\rm o}(x)$ have the forms of 
\begin{equation}
g_{\rm e}(x)\sim \alpha^{(-)}_{\rm e}{\rm e}^{2|x|}
+\beta^{(-)}_{\rm e}{\rm e}^{-2|x|}
\end{equation}
and 
\begin{equation}
g_{\rm o}(x)\sim \left(\alpha^{(-)}_{\rm o}{\rm e}^{2|x|}
+\beta^{(-)}_{\rm o}{\rm e}^{-2|x|}\right){\rm sgn}(x)
\end{equation}
respectively. Generally, the coefficients $\alpha^{(-)}_{\rm e,o}$ of the exponentially growing function are not zero. In (\ref{eq: G-zero-l-c}), on the other hand, those exponentially growing terms should be absent. It is possible only when $B_{\rm e}=B_{\rm o}=0$ and hence 
\begin{equation}
G^{(0)}(x)=0
\label{eq: G-0-zero}
\end{equation}
follows. 
$S^{(1)}(x)$ satisfies 
\begin{equation}
\hat{h}S^{(1)}(x)=G^{(0)}(x)=0
\end{equation}
and hence can be written as
\begin{equation}
A^{(1)}_{\rm e}s_{\rm e}(x)+A^{(1)}_{\rm o}s_{\rm o}(x).
\end{equation}
With use of it, $S(x;\varepsilon)$ is written as
\begin{equation}
S(x;\varepsilon)=A_{\rm e}(\varepsilon)s_{\rm e}(x)+A_{\rm o}(\varepsilon)s_{\rm o}(x)+{\cal O}(\varepsilon^2)
\end{equation}
with $A_{\rm e,o}(\varepsilon)=A^{(0)}_{\rm e,o}+\varepsilon A^{(1)}_{\rm e,o}$. $S(x;\varepsilon)$ becomes
\begin{equation}
A_{\rm e}(\varepsilon)\alpha_{\rm e}-A_{\rm o}(\varepsilon)\alpha_{\rm o}
+
(-A_{\rm e}(\varepsilon)\beta_{\rm e}+A_{\rm o}(\varepsilon)\beta_{\rm o})x+{\cal O}(\varepsilon^2)
\label{eq: asym-AA-left}
\end{equation}
at $x\ll -1$, 
and 
\begin{equation}
A_{\rm e}(\varepsilon)\alpha_{\rm e}+A_{\rm o}(\varepsilon)\alpha_{\rm o}
+
(A_{\rm e}(\varepsilon)\beta_{\rm e}+A_{\rm o}(\varepsilon)\beta_{\rm o})x+{\cal O}(\varepsilon^2)
\label{eq: asym-AA-right}
\end{equation}
at $x\gg 1$. 

From (\ref{eq: asym-left}), on the other hand, we obtain
\begin{equation}
S(x;\varepsilon)\sim (1+\theta_+(\varepsilon))(e^{{\rm i}k_\parallel x}+r(\varepsilon)e^{-{\rm i}k_\parallel x})
\end{equation}
for $x\ll -1$ and 
\begin{equation}
S(x;\varepsilon)\sim (1+\theta_+(\varepsilon))t(\varepsilon) e^{{\rm i}k_\parallel x}
\end{equation}
for $x\gg 1$. 
With use of $k_\parallel=\varepsilon\cos\phi+{\cal O}(\varepsilon^3)$, $\kappa=2+{\cal O}(\varepsilon^2)$ and $1+\theta_+(\varepsilon)=2-\varepsilon+{\cal O}(\varepsilon^2)$, we obtain 
\begin{equation}
S(x;\varepsilon)\sim (2-\varepsilon)\left(1+r(\varepsilon)+
\left(1-r(\varepsilon)\right){\rm i}\tilde{\varepsilon} x \right) +{\cal O}(\varepsilon^2)
\label{eq: asym-r-left}
\end{equation}
for $-1/\tilde{\varepsilon}\ll x\ll -1$
and 
\begin{equation}
S(x;\varepsilon)\sim (2-\varepsilon)t(\varepsilon)
\left(1+{\rm i}\tilde{\varepsilon} x \right) +{\cal O}(\varepsilon^2)
\label{eq: asym-t-right}
\end{equation}
for $1\ll x\ll 1/\tilde{\varepsilon}$. Here $\tilde{\varepsilon}$ denotes $\varepsilon \cos\phi$. Equating (\ref{eq: asym-AA-left}) with (\ref{eq: asym-r-left}), we obtain 
\begin{equation}
\tilde{A}_{\rm e}(\varepsilon)\alpha_{\rm e}-\tilde{A}_{\rm o}(\varepsilon)\alpha_{\rm o}=1+r(\varepsilon)
\label{eq: determine-1}
\end{equation}
and
\begin{equation}
-\tilde{A}_{\rm e}(\varepsilon)\beta_{\rm e}+\tilde{A}_{\rm o}(\varepsilon)\beta_{\rm o}=(1-r(\varepsilon)){\rm i}\tilde{\varepsilon},
\label{eq: determine-2}
\end{equation}
with $\tilde{A}_{\rm e,o}(\varepsilon)=A_{\rm e,o}(\varepsilon)/(2-\varepsilon)$. 
Equating (\ref{eq: asym-AA-right}) with (\ref{eq: asym-t-right}), we obtain 
\begin{equation}
\tilde{A}_{\rm e}(\varepsilon)\alpha_{\rm e}+\tilde{A}_{\rm o}(\varepsilon)\alpha_{\rm o}=t(\varepsilon)
\label{eq: determine-3}
\end{equation}
and
\begin{equation}
\tilde{A}_{\rm e}(\varepsilon)\beta_{\rm e}+\tilde{A}_{\rm o}(\varepsilon)\beta_{\rm o}={\rm i}\tilde{\varepsilon} t(\varepsilon).
\label{eq: determine-4}
\end{equation}
The four equations (\ref{eq: determine-1}), (\ref{eq: determine-2}), (\ref{eq: determine-3}) and (\ref{eq: determine-4}) determine $\tilde{A}_{\rm e,o}(\varepsilon)$, $r(\varepsilon)$ and $t(\varepsilon)$ uniquely. Consequently, we obtain 
\begin{equation}
t(\varepsilon)=\frac{{\rm i}\tilde{\varepsilon} (\alpha_{\rm o}\beta_{\rm e}-\alpha_{\rm e}\beta_{\rm o})}{(\beta_{\rm e}-{\rm i}\tilde{\varepsilon} \alpha_{\rm e})(\beta_{\rm o}-{\rm i}\tilde{\varepsilon} \alpha_{\rm o})}+{\cal O}(\varepsilon^2). 
\end{equation}
Setting $\beta_{\rm e}=0$ (\ref{eq: beta-e-zero}), we arrive at 
\begin{equation}
t(\varepsilon)=\left(1-{\rm i}\tilde{\varepsilon}\frac{\alpha_{\rm o}}{\beta_{\rm o}}\right)^{-1}+{\cal O}(\varepsilon^2)=1+{\rm i}\varepsilon\cos\phi \frac{\alpha_{\rm o}}{\beta_{\rm o}}+{\cal O}(\varepsilon^2),
\label{eq: t-varepsilon}
\end{equation}
i.e. the perfect transmission $$\lim_{\varepsilon \rightarrow 0}t(\varepsilon)=1$$
at low energy limit. From (\ref{eq: t-varepsilon}), we obtain the phase shift $\delta(\varepsilon)$ 
\begin{equation}
\delta(\varepsilon)=\varepsilon\cos\phi\frac{\alpha_{\rm o}}{\beta_{\rm o}}+{\cal O}(\varepsilon^2),
\label{eq: delta-varepsilon}
\end{equation}
which is defined as $(t(\varepsilon)=|t(\varepsilon)|\exp({\rm i}\delta(\varepsilon)))$. The energy dependence of phase shift (\ref{eq: delta-varepsilon}) is consistent with the earlier result\cite{Danshita}. 

The anomalous tunneling is a direct result of (\ref{eq: beta-e-zero}), which follows from the relation (\ref{eq: s-0-e-Psi}) or the fact that the even part $s_{\rm e}(x)$ of the wave function of excitation at low energy limit coincides with the wave function of condensate. For the anomalous tunneling, it is  crucial that the low energy excitation is obtained by infinitesimal deformation of the ground state. This applies to the present case; the low energy excitations are the Goldstone-mode as a consequence of U(1) symmetry breaking.

With this finding, we discuss the anomalous tunneling at finite temperatures. 
Among several theories for Bose system at finite temperatures, we take the Popov approximation\cite{pethicksmith} as an example. From now on, we recover the conventional unit from the dimensionless one. In the Popov approximation, the condensate wave function $\Psi(x)$ satisfies
\begin{equation}
\hat{H_{\rm p}}\Psi(x)=0
\label{eq: GP-popov}
\end{equation}
with $$\hat{H_{\rm p}}=-\frac{\hbar^2\nabla^2}{2m}+V(x)-\mu+U|\Psi(x)|^2+2U\rho(x).$$
The wave function of excitation satisfies 
  \begin{eqnarray}
  &&\left(
  \begin{array}{cc}
  \hat{H}'_{\rm p}& -U \Psi^2(x)\\
  U\Psi^{*2}(x) &-\hat{H}'_{\rm p}\\
  \end{array} 
  \right)
  \left(
  \begin{array}{c}
  u(\textrm{\boldmath $r$};\varepsilon) \\
  v(\textrm{\boldmath $r$};\varepsilon) \\
  \end{array} 
  \right)\nonumber\\
  &&=\varepsilon
  \left(
  \begin{array}{c}
  u(\textrm{\boldmath $r$};\varepsilon) \\
  v(\textrm{\boldmath $r$};\varepsilon) \\
  \end{array} 
  \right)
  \label{eq: Beq-popov}
  \end{eqnarray}
  with $\hat{H}'_{\rm p}=\hat{H}_{\rm p}+U|\Psi(x)|^2$.
  The function $\rho(x)$ is given by
  \begin{equation}
  \rho(x)=\sum_{i}\left(|u_i(\textrm{\boldmath $r$})|^2 f(\varepsilon_i) +|v_i(\textrm{\boldmath $r$})|^2 (1+f(\varepsilon_i))\right).
  \label{eq: rho-def}
  \end{equation}
  Here $i$ is the index of the eigenstate for (\ref{eq: Beq-popov}) under a certain boundary condition and $u_i(\textrm{\boldmath $r$})$ and $v_i(\textrm{\boldmath $r$})$ are the wave functions of the $i$-th eigenstate with energy $\varepsilon_i$ and are normalized by
  $$
  \int{\rm d}\textrm{\boldmath $r$}\left(|u_i(\textrm{\boldmath $r$})|^2-|v_i(\textrm{\boldmath $r$})|^2\right)=1.
  $$ 
  The function $f(\varepsilon)$ is given by  the Planck distribution function $(\exp[\varepsilon/(k_{\rm B}T)]-1)^{-1}$ at temperature $T$. Note that in the presence of $V(x)$ and $\Psi(x)$, $|u_i(\textrm{\boldmath $r$})|^2$ and $|v_i(\textrm{\boldmath $r$})|^2$ in the right-hand side of (\ref{eq: rho-def}) depend only on $x$.
  From (\ref{eq: Beq-popov}), we see that 
\begin{equation}
S(x)=\lim_{\varepsilon\rightarrow 0}(u(\textrm{\boldmath $r$};\varepsilon)+v(\textrm{\boldmath $r$};\varepsilon))
\label{eq: S-popov}
\end{equation}
 satisfies (\ref{eq: GP-popov}) and the even part of (\ref{eq: S-popov}) is proportional to $\Psi(x)$. Following the argument at zero temperature, we conclude that anomalous tunneling occurs at finite temperatures within the Popov approximation. Indeed, we have confirmed numerically that anomalous tunneling occurs for $V(x)\propto \exp(-x^2)$ at finite temperatures within the Popov approximation\cite{Nishiwaki-Kato}. We expect that the anomalous tunneling occurs at finite temperatures even when other approximate theories are taken {\it if} low energy excitations are infinitesimal deformations of the ground state.  

So far we have considered the tunneling by scatterers with the wall geometry. Lastly, we remark on the scattering of excitations by scatterers with other geometries\cite{Fujita,FujitaKato}. Applying the same analysis used so far, we obtain the cross section $\sigma(\varepsilon)$ which depends on energy as
\begin{equation}
\sigma(\varepsilon)\propto \varepsilon^ 3\sim k^3,
\label{eq: Rayleigh2D}
\end{equation}
at low energy for the scattering of excitation by a cylindrical potential and
\begin{equation}
\sigma(\varepsilon)\propto \varepsilon^ 4\sim k^4,
\label{eq: Rayleigh}
\end{equation}
for the scattering of excitation by a spherically symmetric potential. 
(\ref{eq: Rayleigh2D}) and (\ref{eq: Rayleigh}) result from the fact that the wave function of excitations in s-wave channel at low energy coincides with that of condensate. Note that (\ref{eq: Rayleigh}) is consistent with the Rayleigh scattering of sound wave in classical wave mechanics\cite{Landau}. The method presented in this Letter is useful to discuss low energy properties of tunneling and scattering problems by scatterers with various geometries in condensed bose systems. 

In conclusion, we have shown that the anomalous tunneling occurs for symmetric potential barrier $V(x)=V(-x)$. The proof for the occurence of the anomalous tunneling over asymmetric potential barrier $V(x)\ne V(-x)$ is left as a future problem. Another open issue is the origin of anomalous tunneling in the presence of a condensate current \cite{DYK3}. 

We thank D. Yoshioka, S. Watabe, M. Oshikawa, M. Mine, Y. Yamanaka and S. Kurihara for helpful discussions. This work is supported by a Grant-in-Aid for Scientific Research (C) No. 17540314 from the Japan Society for the Promotion of Science.

\end{document}